\documentclass[sigconf]{acmart}
\AtBeginDocument{%
  }

\usepackage{colortbl}
\usepackage{xcolor}
\usepackage{booktabs}
\usepackage{multirow}
\usepackage{booktabs}
\usepackage{array}
\usepackage{makecell}
\usepackage{placeins}
\usepackage{listings}
\usepackage{tcolorbox} 
\tcbset{ colback=gray!15, colframe=white, boxrule=0pt, arc=2pt, left=6pt, right=6pt, top=4pt, bottom=4pt, before skip=8pt,after skip=8pt }

\usepackage{minted} 

\definecolor{codebg}{RGB}{230,230,230}

\usepackage{minted}
\setminted{
    breaklines=true,
    breakanywhere=true,
    fontsize=\small,
    bgcolor=codebg,
    frame=none,
    linenos
}

\newcommand{\phead}[1]{\noindent {\bf #1}}
\newcommand{\uhead}[1]{\noindent\textit{#1}}
\newcommand{\uheadu}[1]{\noindent\underline{\textit{#1}}}

\usepackage{xfp} 
\newcommand{\improve}[2]{\cellcolor{green!\fpeval{min(#2,60)}}#1} 
\newcommand{\regress}[2]{\cellcolor{red!\fpeval{min(#2,60)}}#1} 
\newcommand{\neutral}[1]{#1} 

\setcopyright{acmlicensed}
\copyrightyear{2018}
\acmYear{2018}
\acmDOI{XXXXXXX.XXXXXXX}
\acmConference[Conference acronym 'XX]{Make sure to enter the correct
  conference title from your rights confirmation email}{June 03--05,
  2018}{Woodstock, NY}
\acmISBN{978-1-4503-XXXX-X/2018/06}

\begin{document}

\title{Evaluating Software Process Models for Multi-Agent Class-Level Code Generation}


\author{Wasique Islam Shafin}
\affiliation{%
  \institution{SPEAR Lab, Concordia University}
  \city{Montreal, QC}
  \country{Canada}
}
\email{w_shafin@encs.concordia.ca}

\author{Md Nakhla Rafi}
\orcid{0009-0005-4707-8985}
\affiliation{%
  \institution{SPEAR Lab, Concordia University}
  \city{Montreal, QC}
  \country{Canada}
}
\email{r_mdnakh@encs.concordia.ca}

\author{Zhenhao Li}
\orcid{0000-0002-4909-1535}
\affiliation{%
  \institution{York University}
  \city{Toronto, Ontario}
  \country{Canada}
}
\email{lzhenhao@yorku.ca}

\author{Tse-Hsun (Peter) Chen}
\orcid{0000-0003-4027-0905}
\affiliation{%
  \institution{SPEAR Lab, Concordia University}
  \city{Montreal, QC}
  \country{Canada}
}
\email{peterc@encs.concordia.ca}

\renewcommand{\shortauthors}{Shafin et al.}

\begin{abstract}
Modern software systems require code that is not only functional but also maintainable and well-structured. Although Large Language Models (LLMs) are increasingly used to automate software development, most studies focus on isolated, single-agent function-level generation. This work examines how process structure and role specialization shape multi-agent LLM workflows for class-level code generation. We simulate a Waterfall-style development cycle covering Requirement, Design, Implementation, and Testing using three LLMs (GPT-4o-mini, DeepSeek-Chat, and Claude-3.5-Haiku) on 100 Python tasks from the ClassEval benchmark. Our findings show that multi-agent workflows reorganize, rather than consistently enhance, model performance. Waterfall-style collaboration produces cleaner and more maintainable code but often reduces functional correctness (–37.8\% for GPT-4o-mini and –39.8\% for DeepSeek-Chat), with Claude-3.5-Haiku as a notable exception (+9.5\%). Importantly, process constraints shift failure characteristics: structural issues such as missing code decrease, while semantic and validation errors become more frequent. Among all stages, Testing exerts the strongest influence by improving verification coverage but also introducing new reasoning failures, whereas Requirement and Design have comparatively modest effects. Overall, this study provides empirical evidence that software process structure fundamentally alters how LLMs reason, collaborate, and fail, revealing inherent trade-offs between rigid workflow discipline and flexible problem-solving in multi-agent code generation.
\end{abstract}




\maketitle
\section{Introduction}
\label{sec:intro}
Large Language Models (LLMs) have become increasingly influential in Software Engineering (SE), assisting in diverse tasks such as code generation \cite{Liu02, mu2023clarifygpt, lin2024soen, hong2024metagpt, qian2023chatdev, chen2024code}, code completion and assistance \cite{github_copilot_2025, windsurf_2025, tabnine_2025, cursor_2025}, fault localization and testing \cite{kang2024quantitative, rafi2024impact, wang2024software}, and automated program repair \cite{berabi2024deepcode, zhang2024autocoderover, Bouzenia01, Yang01, xia2024agentless, Yin-01}. General-purpose models such as ChatGPT \cite{openai_chatgpt_2025}, Claude \cite{anthropic_claude_2025}, and DeepSeek \cite{deepseek_chat_2025} have further extended the reach of LLMs into broader SE workflows, including debugging, documentation, and architecture planning \cite{zhang2023unifying, zheng2023survey, jiang2024survey}.  As software systems continue to grow in complexity, the ability of LLMs to generate maintainable, reliable, and secure code has become central to their practical deployment in SE \cite{dong2025survey, Alharbi2025}.

Early advances in automated code generation primarily focused on function-level synthesis, where models such as Codex \cite{chen2021evaluating}, AlphaCode \cite{Li2022competition}, and CodeT5+ \cite{wang2023codet5} demonstrated strong capabilities in producing short, self-contained functions. This line of research evolved through increasingly complex benchmarks like HumanEval \cite{chen2021evaluating}, MBPP \cite{austin2021program}, and BigCodeBench \cite{zhuo2024bigcodebench}, which standardize evaluations using metrics such as Pass@k. More recent frameworks, like Chain-of-Programming \cite{hou2024chain}, interactive test-driven generation \cite{fakhoury2024llm}, and reasoning-enhanced prompting \cite{li2025structured, wei2022chain}, extend these capabilities toward more realistic and process-aware programming tasks. However, while these methods improve functional correctness, most evaluations still consider isolated functions rather than interconnected classes that reflect real-world software modularity and maintainability concerns.

To address this limitation, class-level benchmarks like \textit{ClassEval} \cite{du2023classeval} have emerged to capture object-oriented code generation, assessing how well models handle multi-method dependencies and shared class state. Subsequent extensions such as Xue et al.~\cite{xue2025classeval} and Chen et al.~\cite{chen2024reasoning} expand ClassEval in multiple languages and integrate error analysis frameworks, offering deeper insight into reasoning-based code generation. However, despite these advances, the majority of research remains focused on single-LLM paradigms. At the same time, multi-agent frameworks such as ChatDev \cite{qian2023chatdev}, MetaGPT \cite{hong2024metagpt}, SOEN-101 \cite{lin2024soen}, and AgileCoder \cite{nguyen2025agilecoder} have introduced collaborative, role-based approaches that emulate human software processes by assigning distinct agents to roles such as Product Manager, Architect, Developer, and Tester. These systems have shown improvements in modularity, reasoning, and error mitigation at the function level. However, evaluations have not yet extended to class-level tasks, leaving an open question: 
\textit{Do multi-agent LLM workflows enhance class-level code quality and correctness, and which development activities matter most?}

In this paper, we bridge these two research directions by studying how a multi-agent workflow can operate within a classical Waterfall software process model for class-level code generation using \textit{ClassEval}. We use Waterfall because its sequential structure (Requirement → Design → Implementation → Testing) makes it easier to study the effect of each development activity. Our study assigns distinct roles (Requirement Engineer, Architect, Developer, and Tester) to specialized LLM agents, allowing us to investigate how each stage of the Waterfall process contributes to the final implementation. Through an ablation-based experimental design, we evaluate three LLM backends (GPT-4o-mini, DeepSeek-Chat, Claude-3.5-Haiku) and compare them against direct prompting as the baseline, measuring (1) functional correctness (pass@1), (2) software quality and maintainability using SonarQube metrics, (3) identify frequently encountered error types, and (4) detailed failure categories through a taxonomy-driven error analysis.

Our results show that multi-agent LLM workflows do not uniformly enhance class-level code correctness but significantly affect code quality and error characteristics. Waterfall-style workflows give cleaner, more maintainable code; however, pass@1 accuracy declines for gpt-4o-mini (-37.8\%) and deepseek-chat (-39.8\%), while claude-3.5-haiku improves (+9.5\%), indicating model-dependent effects. These workflows also increase runtime errors by 10-53\%, with ValueError, AssertionError, and TypeError occurring most frequently. Although structural issues such as \textit{Missing Code} are reduced, \textit{Semantic Failure} and \textit{Return Mismatches} become more common. Among development activities, Testing has the strongest influence: including it improves verification but increases reasoning and validation errors; excluding it yields the poorest correctness yet highest code quality. In contrast, the Requirements and Design stages have minimal impact on correctness or error types. The complete Waterfall configuration produces the lowest correctness and greatest adaptability concerns, suggesting that over-structuring the workflow constrains the flexibility of the final code. Incorporating these development activities into a multi-agent workflow enforces stricter validation and verification, but this also makes the code more error-prone and reduces its overall adaptability, intentionality, and correctness.

Overall, this paper makes the following contributions:
\begin{itemize}
    \item We propose the first systematic study of multi-agent LLM workflows applied to class-level code generation using ClassEval in the form of a Waterfall process model.
    
    \item We perform an empirical comparison of Waterfall-activity ablations that isolates the effect of every development activity on functional correctness and code quality.

    \item We found that multi-agent LLM workflows do not consistently improve class-level code correctness but do significantly influence code quality; testing has the greatest impact, strengthening verification while also increasing reasoning errors, whereas overly structured Waterfall workflows reduce both adaptability and correctness.

    \item We make all code and data from our experiments publicly available to enable reproduction, verification, and further research. \cite{anonymous2025repository}
\end{itemize}

\section{Background and Related Work}
LLMs have rapidly advanced the field of automated code generation, with growing research spanning model development, benchmarking, and multi-agent collaboration frameworks. This section first reviews the evolution of LLM-based code generation, then discusses recent progress in agent-driven workflows that simulate software engineering processes. We next summarize existing benchmarks and surveys, particularly those addressing class-level generation tasks. Finally, we identify the research gap motivating our study, which evaluates class-level code generation within structured multi-agent workflows, a dimension that remains missing in current literature.

\subsection{LLM in Code Generation}
The evolution of LLMs for code generation has developed rapidly since 2021, beginning with Codex \cite{chen2021evaluating}, which demonstrated that models fine-tuned on large-scale code bases can translate natural language into executable code. Subsequent models, such as AlphaCode \cite{Li2022competition} and CodeT5+ \cite{wang2023codet5}, expanded further by handling competitive programming challenges and refining generation through encoder–decoder architectures and feedback-based ranking mechanisms. By 2023–2024, works like PANGU-CODER2 \cite{shen2023pangu} and query-aligned prompting methods \cite{ma2023query, li2025structured, wei2022chain} improved model reasoning and task alignment. More recent research has shifted toward embedding LLMs within full software workflows, incorporating test-driven development and interactive debugging \cite{hou2024chain, fakhoury2024llm}. Surveys published in 2025 \cite{dong2025survey, Alharbi2025} highlight a transition from code snippet generation to end-to-end software development workflows, encompassing planning, optimization, and integration.

\subsection{Multi-Agent Code Generation Frameworks}
In parallel, multi-agent frameworks have emerged to emulate collaborative software engineering processes using role-specialized LLMs. Frameworks like ChatDev \cite{qian2023chatdev} and MetaGPT \cite{hong2024metagpt} assign agents roles such as Product Manager, Developer, and Tester, coordinating through dialogue (chatting) to design, implement, and test software solutions. These systems primarily operate at the function level, evaluating outcomes via completeness, executability, and pass@k metrics. SOEN-101 \cite{lin2024soen} extends this approach by structuring multi-agent workflows according to software process models (e.g., Waterfall, TDD, Scrum), while MAS \cite{cemri2025multi} categorizes common failure types across frameworks, such as system design errors and inter-agent misalignments. AgileCoder \cite{nguyen2025agilecoder} further introduces Agile-inspired iterative sprints, highlighting adaptability and incremental refinement. Collectively, these studies confirm that agent-based LLMs improve task division and communication in software generation, but they remain confined to function-level code, leaving class-level software development largely unexplored.

\subsection{Benchmarks and Surveys}
Benchmarking remains central to evaluating progress in LLM-based code generation. Widely adopted datasets such as HumanEval \cite{chen2021evaluating}, MBPP \cite{austin2021program}, and BigCodeBench \cite{zhuo2024bigcodebench} measure correctness using the pass@k metric, focusing on isolated, function-level tasks. Surveys by Jiang et al. \cite{jiang2024survey}, Zhang et al. \cite{zhang2023unifying}, and Zheng et al. \cite{zheng2023survey} systematically compare LLM performance across these benchmarks, revealing ongoing challenges in reasoning, building complex code, and error propagation.

Recognizing the limitations of function-level assessments, ClassEval \cite{du2023classeval} was introduced to evaluate class-level generation tasks involving interdependent methods, shared variables, and internal class logic. Follow-up works such as Xue et al. \cite{xue2025classeval} expanded ClassEval to Java and C++, incorporating new metrics and error analysis, while Chen et al. \cite{chen2024reasoning} examined reasoning performance across HumanEval and ClassEval. Complementary frameworks like CoCoST \cite{he2024cocost} integrate automated test generation and web search to assess broader problem-solving capabilities. These advances collectively provide a foundation for studying class-level code generation, yet none explore how structured, agent-driven workflows like waterfall influence performance on such complex tasks.

\subsection{Limitations and Research Gap}
While multi-agent frameworks have proven effective for function-level collaboration and class-level benchmarks have captured higher program complexity, these two directions remain largely disconnected. Existing literature does not examine how structured agent workflows affect class-level code generation quality, nor how different software development activities (requirements analysis, design, implementation, and testing) contribute to correctness and maintainability.
Our study bridges this gap by evaluating class-level code generation through a Waterfall-inspired multi-agent workflow. This structured setup enables activity-wise analysis of LLM behaviour by tracing how each development activity impacts the final code quality across models. In contrast to prior work emphasizing model architecture or function-level benchmarks, our contribution lies in characterizing how collaborative process design has an impact on what LLMs produce, offering new insights into class-level code generation that is aware of the software engineering process.

\section{Methodology and Experiment Setup}
To understand how different development activities influence the quality of code produced during \textit{class-level} code generation, we use agents to emulate the \textbf{Waterfall} model. Its straightforward and sequential nature allows us to examine how individual development activities affect the resulting code. 
We follow a step-by-step linear process, where development flows downwards through distinct phases (from requirements to design, implementation, and testing). 
\begin{figure}[h]
    \centering
    \includegraphics[width=0.95\columnwidth]{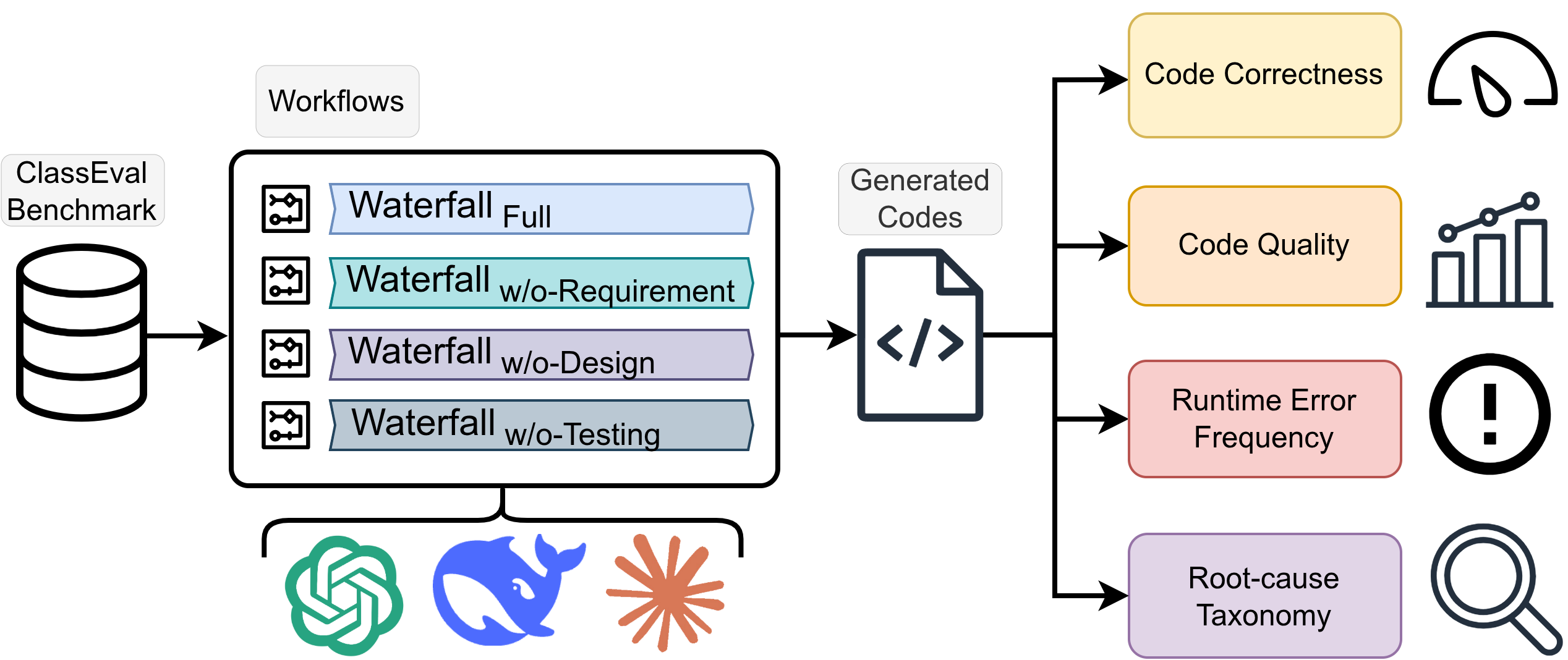}
    \caption{Approach Overview}
    \Description{Approach Overview}
    \label{fig:overview_workflow}
\end{figure}
Figure~\ref{fig:overview_workflow} provides an overview of our study design. 
Our experiment design is organized around two main components:
(1) \textbf{Agent Roles}, which define the LLM agents and their responsibilities, and
(2) \textbf{Agent Communication}, which describes how these agents interact throughout the development phases under the Waterfall model.

\subsection{Agent Roles}

\begin{table*}[h]
\centering
\scriptsize
\caption{Tasks, instructions, and context associated with each role prompt}
\label{tab:prompt_construct}
\resizebox{\textwidth}{!}{
\begin{tabular}{p{2cm}|p{4cm} p{5cm} p{3cm}}
\hline
\textbf{Role} & \textbf{Task} & \textbf{Instructions} & \textbf{Context}\\ \hline
Requirement Engineer & Analyze the task description and write a requirements document. & 1) Review the task description. 2) Write a requirements document that accounts for method signatures and class variables. & Task description (user requirement). \\ \hline
Architect & Design the high-level components and structure of the code. & 1) Review the provided documents. 2) Write a design document that preserves method definitions and serves as a guide for developers. & Requirements document. \\ \hline
Developer & Implement Python code that meets all requirements. & 1) Review the provided documents. 2) Write clean, efficient, and readable code following best practices. Use a single class structure and keep method definitions unchanged. & Design document. \\
& Fix code while ensuring all requirements are met. & 1) Review the test reports and provided documents. 2) Revise the code to make it efficient, readable, and consistent with best practices. & Test report and original code. \\ \hline
Tester & Design test cases that verify all requirements. & 1) Review the provided documents. 2) Write test cases that preserve method definitions, cover normal, edge, and error conditions, and include at least five test cases per method. & Task description (user requirement). \\ 
& Write a Python test script using the \texttt{unittest} framework. & 1) Review the provided documents. 2) Write test scripts with one test case for each scenario, reusing existing modules where possible. & Test case document. \\ 
& Write a test failure report. & 1) Review the test execution results. 2) Document the outcomes in a test report. & Test execution results. \\ \hline
\end{tabular}
}
\end{table*}
We employ four LLM agents to represent the primary activities of the software development life cycle (SDLC) within a \textit{Waterfall} process: \textit{requirements}, \textit{design}, \textit{implementation}, and \textit{testing}. 
These agents are (1) \textbf{Requirement Engineer}, (2) \textbf{Architect}, (3) \textbf{Developer}, and (4) \textbf{Tester}, each focusing on a specific development activity and their interactions, mirror their real-life counterparts. 

Every agent uses the same unified prompt template, with differences in role-specific details. 
We adapt and modify the prompt structure from SOEN-101~\cite{lin2024soen} to enable class-level generation workflows, whereas it focused only on function-level generation only. The prompt template defines how each agent interprets its assigned role, consumes relevant context, and generates the corresponding software artifact. A simplified example of the template is shown below: 


\begin{small}
\begin{verbatim}
{
  "Role": "You are a [role] delegated for [task]",
  "Instruction": "According to the Context, 
                  [role-specific-instruction]",
  "Example": "[document example]",
  "Context": "[context]",
  "Question": "Follow the instructions. 
               The [document] must satisfy the requirements."
}
\end{verbatim}
\end{small}

\uhead{role} defines the agent's role in the process model (e.g., Developer or Tester),  
\uhead{task} specifies the objective of that role, 
\uhead{document} represents the artifact to be produced,  
\uhead{example} provides a reference format,  
\uhead{role-specific-instruction} offers detailed guidance tailored to the task, and  
\uhead{context} supplies necessary inputs or artifacts from earlier stages.  

Table~\ref{tab:prompt_construct} summarizes the \textit{tasks}, \textit{instructions}, and \textit{contexts} assigned to each agent, showing how they collaborate to generate, refine, and verify software artifacts.
By delegating well-defined tasks to specialized agents, our approach enables systematic experimentation within a Waterfall workflow, refining how agents collaborate and exchange information across sequential phases.

\subsection{Agent Communication}
For our study, the agent communication is based on the Waterfall process model. This process divides work into a sequence of well-defined activities (e.g., requirements, design, implementation, and testing) where each activity depends on the deliverables of the previous one. Within this process, the agents with specialized roles collaborate across the sequential stages with formal handoffs and documentations to produce a final product. In addition, we designed the agents to provide feedback for self-refinement \cite{AmanMadaan01CMU}.

This section describes how agents coordinate within these structured activities and how their interactions are organized into a complete Waterfall-inspired workflow.

\subsubsection{Communication}
\begin{figure*}[t]
    \centering
    \includegraphics[width=1\textwidth]{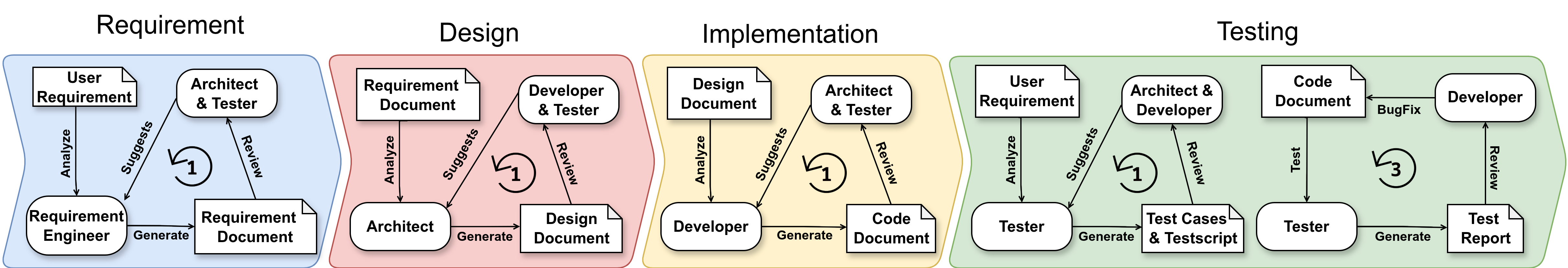}
    \caption{Activities of Waterfall Workflow}
    \Description{Activities of Waterfall Workflow}
    \label{fig:phases}
\end{figure*}

The agent communication process is divided into four sequential activities: \textit{Requirement}, \textit{Design}, \textit{Implementation}, and \textit{Testing}, corresponding directly to the core stages of the \textit{Waterfall} software development life cycle (SDLC). As illustrated in Figure~\ref{fig:phases}, these four activities collectively form a complete Waterfall-inspired communication process among the LLM agents.

\uheadu{Requirement Activity.}  
The \textit{Requirement Engineer} agent begins the process by transforming the \texttt{User Requirement} into a formal \texttt{Requirement Document} that captures the high-level functional and structural details, such as method signatures, class variables, and inter-dependencies.  
Communication in this phase primarily involves clarification and validation of user needs by the agent. The document is self-refined \cite{AmanMadaan01CMU} through feedback from the \textit{Architect} and \textit{Tester} to ensure fulfillment of acceptance criteria before progressing to the next phase.

\uheadu{Design Activity.}  
The \textit{Architect} agent receives the completed \texttt{Requir}- texttt{ement Document} and produces a \texttt{Design Document} that outlines the system's architecture, class structure, and method interactions. 
This phase formalizes the blueprint for development, and feedback from the \textit{Developer} and \textit{Tester} is incorporated \cite{AmanMadaan01CMU} only to resolve inconsistencies or design ambiguities. This phase reflects Waterfall's controlled, document-driven communication before implementation begins.

\uheadu{Implementation Activity.}  
The \textit{Developer} agent implements the \texttt{Desi}- \texttt{gn Document} by producing a \texttt{Code Document} that the agent realizes the required functionality in executable code using a single class structure. 
Like that of previous phases, our framework allows self-refinement \cite{AmanMadaan01CMU} based on suggestions from the \textit{Architect} and \textit{Tester} to correct implementation deviations or improve maintainability. This controlled interaction simulates real-world review cycles that occur before code is formally handed over for testing.

\uheadu{Testing Activity.}  
The \textit{Tester} agent derives \texttt{Test Cases} directly from the \texttt{User Requirement} and translates them into executable \texttt{Test Scripts}.  
These test cases are self-refined \cite{AmanMadaan01CMU} using feedback from the \textit{Architect} and \textit{Developer} to resolve ambiguity and ensure accurate test coverage.  
The \textit{Tester} executes the \texttt{Test Script} against the \texttt{Code Document}, producing a \texttt{Test Report} that summarizes the detected faults and performance outcomes. Finally, the \textit{Developer} may perform up to three bug-fix cycles based on this report, after which the process concludes. Thus, reflecting the validation and closure phase of the Waterfall cycle.

\subsubsection{Studying the Effect of Each Development Activity}
To understand the impact of each activity, we take the complete Waterfall and remove one activity at a time, and then observe how its absence affects overall performance, code quality, and the possible issues. This ablation-based approach allows us to identify which phases most strongly affect the correctness, completeness, and reliability of the final output. We retain the \texttt{Implementation} phase across all the configurations, as code generation is crucial to producing the executable artifact required for evaluation.
For our study, we have 4 configurations of Waterfall, where \textit{w/o} denotes the removal of the corresponding activity:
1) \textbf{Waterfall$_{Full}$}, 
2) \textbf{Waterfall$_{w/o Requirement}$},
3) \textbf{Waterfall$_{w/o Design}$},
4) \textbf{Waterfall $_{w/o Testing}$}.

\subsection{Dataset}
We use the \textbf{ClassEval} dataset~\cite{du2023classeval}, a human-curated benchmark for evaluating LLMs on \textit{class-level Python coding tasks}. Unlike prior benchmarks such as \textbf{HumanEval}~\cite{chen2021evaluating} and \textbf{MBPP} \cite{austin2021program}, which assess short, function-level code generation, \textbf{ClassEval} emphasizes the generation of complete classes that integrate multiple methods, fields, and dependencies. This class structure introduces more reasoning challenges related to design consistency, state management, and inter-method interaction, i.e., features that closely resemble real-world software development than that of isolated coding exercises.
The dataset contains 100 programming assignments, each requiring the implementation of a complete Python class. It was developed over approximately 500 person-hours, emphasizing diversity and realism in software development scenarios.

Each ClassEval task provides:
(1) \uhead{Class Skeleton.} This includes import statements, the class name, a brief description, and the class constructor, along with detailed specifications for each method. Each method specification outlines the method signature, functional description, parameters, and return values, and example input/output. These skeletons serve as the task descriptions in our study.
(2) \uhead{Ground-Truth Implementation.} A complete and correct reference implementation of the target class in Python, serving as the canonical solution. 
(3) \uhead{Test Cases.} An average of 33.1 test cases per class to rigorously assess code correctness. On average, each class comes with 33.1 test cases designed to rigorously evaluate correctness. These tests allow us to measure both function-level and class-level accuracy.

We want to study the impact of each activity on different quality aspects at class-level. 
For our experiments, we adopt the holistic class-generation approach, which generates an entire class at once. 
This strategy has been shown to achieve better performance~\cite{du2023classeval, xue2025classeval}, particularly with the improved context handling capabilities of modern LLMs.

\subsection{LLM Choice and Implementation Details}
We evaluate our workflows using three contemporary LLMs: \texttt{GPT-4o Mini} (gpt-4o-mini-2024-07-18), \texttt{deep seek-chat} (DeepSeek-V3-0324), and \texttt{claude-3.5-haiku} (claude-3-5-haiku-20241022). All three models represent lightweight yet high-performance variants of their respective families, making them suitable for our multi-agent and workflow-based code generation study.
By applying identical workflows across models, we hope to observe whether different LLMs show distinct sensitivities to 
different development activities in the Waterfall process model, and to determine which phases of the Waterfall pipeline influence their performance.
For all models, the temperature is set to 0.8 to balance creativity and determinism during generation.
\section{Results}
In this section, we present the results to our research questions.

\subsection{RQ1: What is the accuracy of the generated code?}
\label{sub:rq1}
\phead{Motivation:}
Class-level code generation differs from method-level generation, as it requires handling higher-level design decisions, inter-method interactions, and overall class organization rather than isolated function logic. 
In this RQ, we study how applying the \textit{Waterfall} process model and enforcing/implementing each of its development activities affect the quality of generated code, and what trade-offs arise in terms of code correctness or quality compared to unstructured direct prompting.
\begin{table*}[h!]
    \centering
    \small
    \caption{Comparison of \textit{gpt-4o-mini}, \textit{deepseek-chat}, and \textit{claude-3.5-haiku} across Waterfall variants. Each cell shows the raw metric and its relative change (\%) from the \textit{RawPrompt} baseline. Green indicates improvement; red indicates regression. Higher is better for \textit{Pass@1}, lower for both \textit{Software Quality} and \textit{Clean Code Attributes}.}
    \begin{tabular}{ll|cc|cc|ccc}
    \hline
    \textbf{Model} & \textbf{Approach} & \multicolumn{2}{c|}{\textbf{Pass@1 (↑)}} & \multicolumn{2}{c|}{\textbf{Software Quality (↓)}} & \multicolumn{3}{c}{\textbf{Clean Code Attribute (↓)}} \\
    \cline{3-9}
     &  & Class & Function & Reliability & Maintainability & Consistency & Intentionality & Adaptability \\
    \hline
    \multirow{5}{*}{\parbox{1.0cm}{\centering gpt-4o \\ -mini}}
    & RawPrompt       & 0.35 & 0.6853 & 0 & 0.2254 & 0.1211 & 0.0908 & 0.0135 \\
    & Waterfall$_{Full}$  & 0.21 (\cellcolor{red!60}-40\%) & 0.5478 (\cellcolor{red!40}-20\%) & 0.0044 \cellcolor{red!20} & 0.2188 (\cellcolor{green!20}-3\%) & 0.0774 (\cellcolor{green!40}-36\%) & 0.0796 (\cellcolor{green!20}-12\%) & 0.0663 (\cellcolor{red!80}+391\%) \\
    & Waterfall$_{w/o-Requirement}$   & 0.23 (\cellcolor{red!60}-34\%) & 0.5458 (\cellcolor{red!40}-20\%) & 0.0023 \cellcolor{red!20} & 0.1991 (\cellcolor{green!40}-12\%) & 0.0801 (\cellcolor{green!40}-34\%) & 0.0618 (\cellcolor{green!40}-32\%) & 0.0595 (\cellcolor{red!80}+341\%) \\
    & Waterfall$_{w/o-Design}$   & 0.21 (\cellcolor{red!60}-40\%) & 0.5896 (\cellcolor{red!20}-14\%) & 0.0072 \cellcolor{red!20} & 0.2141 (\cellcolor{green!20}-5\%) & 0.0818 (\cellcolor{green!40}-33\%) & 0.0987 (\cellcolor{red!20}+9\%) & 0.0409 (\cellcolor{red!80}+203\%) \\
    & Waterfall$_{w/o-Testing}$    & 0.22 (\cellcolor{red!60}-37\%) & 0.5458 (\cellcolor{red!40}-20\%) & 0 & 0.1820 (\cellcolor{green!40}-19\%) & 0.0688 (\cellcolor{green!40}-43\%) & 0.0738 (\cellcolor{green!20}-19\%) & 0.0393 (\cellcolor{red!80}+191\%) \\
    \hline
    \multirow{5}{*}{\parbox{1.0cm}{\centering deepseek \\ -chat}}
    & RawPrompt       & 0.46 & 0.7529 & 0 & 0.2015 & 0.1085 & 0.0558 & 0.0372 \\
    & Waterfall$_{Full}$  & 0.32 (\cellcolor{red!60}-30\%) & 0.6255 (\cellcolor{red!40}-17\%) & 0.0059 \cellcolor{red!20} & 0.2075 (\cellcolor{red!20}+3\%) & 0.0889 (\cellcolor{green!40}-18\%) & 0.0692 (\cellcolor{red!40}+24\%) & 0.0514 (\cellcolor{red!40}+38\%) \\
    & Waterfall$_{w/o-Requirement}$   & 0.29 (\cellcolor{red!60}-37\%) & 0.6554 (\cellcolor{red!20}-13\%) & 0.0109 \cellcolor{red!20} & 0.2049 (\cellcolor{red!20}+2\%) & 0.0916 (\cellcolor{green!40}-16\%) & 0.0632 (\cellcolor{red!20}+13\%) & 0.0567 (\cellcolor{red!40}+52\%) \\
    & Waterfall$_{w/o-Design}$   & 0.31 (\cellcolor{red!60}-33\%) & 0.6793 (\cellcolor{red!20}-10\%) & 0 & 0.2001 (\cellcolor{green!20}-1\%) & 0.0703 (\cellcolor{green!40}-35\%) & 0.0575 (\cellcolor{red!20}+3\%) & 0.0724 (\cellcolor{red!80}+95\%) \\
    & Waterfall$_{w/o-Testing}$    & 0.19 (\cellcolor{red!80}-59\%) & 0.5378 (\cellcolor{red!60}-29\%) & 0.0040 \cellcolor{red!20} & 0.1564 (\cellcolor{green!40}-22\%) & 0.0842 (\cellcolor{green!40}-22\%) & 0.0321 (\cellcolor{green!40}-43\%) & 0.0421 (\cellcolor{red!20}+13\%) \\
    \hline
    \multirow{5}{*}{\parbox{1.0cm}{\centering claude-3 \\ -5-haiku}}
    & RawPrompt       & 0.21 & 0.3685 & 0 & 0.1659 & 0.0869 & 0.0435 & 0.0356 \\
    & Waterfall$_{Full}$  & 0.22 (\cellcolor{green!20}+5\%) & 0.6255 (\cellcolor{green!80}+70\%) & 0 & 0.1736 (\cellcolor{red!20}+5\%) & 0.0825 (\cellcolor{green!20}-5\%) & 0.0282 (\cellcolor{green!40}-35\%) & 0.0629 (\cellcolor{red!80}+77\%) \\
    & Waterfall$_{w/o-Requirement}$   & 0.24 (\cellcolor{green!20}+14\%) & 0.6056 (\cellcolor{green!80}+64\%) & 0.0024 \cellcolor{red!20} & 0.1591 (\cellcolor{green!20}-4\%) & 0.0807 (\cellcolor{green!20}-7\%) & 0.0546 (\cellcolor{red!40}+26\%) & 0.0261 (\cellcolor{green!20}-27\%) \\
    & Waterfall$_{w/o-Design}$   & 0.29 (\cellcolor{green!40}+38\%) & 0.6275 (\cellcolor{green!80}+70\%) & 0 & 0.1734 (\cellcolor{red!20}+5\%) & 0.0759 (\cellcolor{green!20}-13\%) & 0.0477 (\cellcolor{red!20}+10\%) & 0.0499 (\cellcolor{red!40}+40\%) \\
    & Waterfall$_{w/o-Testing}$    & 0.17 (\cellcolor{red!20}-19\%) & 0.4343 (\cellcolor{green!40}+18\%) & 0.0021 \cellcolor{red!20} & 0.1501 (\cellcolor{green!20}-10\%) & 0.0813 (\cellcolor{green!20}-6\%) & 0.0396 (\cellcolor{green!20}-9\%) & 0.0292 (\cellcolor{green!20}-18\%) \\
    \hline
    \end{tabular}
    \label{tab:approach_comparison_all}
    \vspace{2pt}
    \raggedright
    \textbf{Note:} The \textit{Security} and \textit{Responsibility} columns are omitted from the table as all their values were zero.
\end{table*}

\phead{Approach:}
For our class-level study on LLM agent–generated code, we compare the outputs of Waterfall-inspired variants with a baseline \textit{RawPrompt} workflow, where the model generates code directly from task descriptions without any intermediate development stages.
We evaluate the effect of development activities by studying the code quality generated when each activity is removed one at a time. 
We evaluate the generated code based on two aspects: (1) \textbf{code correctness}, measured using the \textit{Pass@1} metric from the ClassEval benchmark, and (2) \textbf{code quality}, analyzed using static analysis with SonarQube.
This combined evaluation allows us to assess both the execution success of generated code and its overall code quality.

\uheadu{Code Correctness.} 
We use \textit{Pass@k} \cite{chen2021evaluating, kulal2019spoc} as the primary measure of code correctness, representing the expected proportion of problems correctly solved given $k$ generated code samples:

\begin{equation}
\text{Pass@}k = \mathbb{E}_{\text{Problems}}\left[1 - \frac{\binom{n - c}{k}}{\binom{n}{k}}\right]
\end{equation}

Here, $n$ denotes the total number of generated samples, $c$ represents the number of correct solutions, and $k$ is the number of evaluated samples. 
We report both \textbf{class-level} and \textbf{function-level} \textit{Pass@1} results. 
At the class level, correctness requires all associated methods in a class to pass their respective tests, whereas at the function level, each method is evaluated independently. 
Focusing on \textit{Pass@1} provides a strict single-attempt measure of correctness, capturing how often the first generated code sample passes all tests. 
All evaluations are conducted using the ClassEval framework.

\uheadu{Code Quality.} 
To complement test-based correctness, we analyze code quality using \textbf{SonarQube}\footnote{\url{https://www.sonarsource.com/products/sonarqube/}}, which detects and categorizes issues into two major groups: \textit{Software Quality} and \textit{Clean Code Attributes}.
The \textit{Software Quality}\footnote{\url{https://docs.sonarsource.com/sonarqube-server/10.6/user-guide/clean-code/software-qualities}} group includes 
(1) \textbf{Security} - Safeguarding software against unauthorized access, use, or damage (protection from misuse), 
(2) \textbf{Reliability} - Ensuring software consistently performs as expected over time (consistent performance), and 
(3) \textbf{Maintainability} - Ease of understanding, fixing, and improving software code (ease of modification).
The \textit{Clean Code}\footnote{\url{https://docs.sonarsource.com/sonarqube-server/10.6/user-guide/clean-code/definition}} group evaluates 
(1) \textbf{Consistency} - Writing code in a uniform and conventional manner (adherence to conventions), 
(2) \textbf{Intentionality} - Ensuring code is precise and purposeful (clarity and completeness), 
(3) \textbf{Adaptability} - Structuring code for easy evolution and confident development (modular and reusable structure), and 
(4) \textbf{Responsibility} - Considering ethical obligations and societal impact in code (lawful and ethical coding behavior).


SonarQube reports issue density normalized per 10 non-comment lines of code (ncLOC), defined as:
\begin{equation}
\text{Issue Density per 10 ncLOC} = \frac{\text{SonarQube Issues}}{\text{Total ncLOC}} \times 10,
\end{equation}

\noindent where \textit{SonarQube Issues} denotes the number of detected problems and \textit{Total ncLOC} refers to the total non-comment lines of code. 
Each workflow's generated code is imported into SonarQube as a separate project, and the resulting metrics are collected from the SonarQube dashboard.
Higher issue density indicates poorer code quality, providing a complementary perspective to the test-based correctness results.

\phead{Results:} 
Table \ref{tab:approach_comparison_all} presents the comparative performance of all models under different Waterfall configurations.

\uhead{Accuracy.}
\textbf{\textit{Pass@1 accuracy declines under Waterfall variants for \emph{gpt-4o-mini} and \emph{deepseek-chat}, whereas \emph{claude-3.5-haiku} benefits from structured prompting.}}
Both \textbf{gpt-4o-mini} and \textbf{deepseek-chat} show a consistent decrease in accuracy when the Waterfall workflow is applied. For these models, the \textit{RawPrompt} approach yields the highest performance (0.35 and 0.46 at class level; 0.6853 and 0.7529 at function level, respectively). Across all Waterfall variants, class-level accuracy declines by roughly 30–40\% and function-level accuracy by 10–25\%. Among the variants, \textit{w/o Requirement} and \textit{w/o Design} retain comparatively better results, while \textit{w/o Testing} produces the weakest performance for both models.
In contrast, \textit{claude-3.5-haiku} exhibits the opposite trend: Waterfall variants generally enhance performance relative to \textit{RawPrompt}. Notably, \textit{w/o Design} achieves the highest overall accuracy (0.29 class, 0.6275 function), and even the full Waterfall workflow provides substantial gains, particularly at the function level (+70\% over baseline).
Interestingly, the results reveal that \textbf{Pass@1 drops significantly from function-level to class-level}, suggesting that it is far easier for a model to generate a single correct function than to produce an entire class where all functions are correct.

\uhead{Software Quality.}
\textbf{\textit{Reliability decreases across all models under Waterfall workflows, while maintainability improves clearly for \emph{gpt-4o-mini} but shows mixed outcomes for \emph{deepseek-chat} and \emph{claude-3.5-haiku}.}}
For \textbf{gpt-4o-mini}, all Waterfall variants lead to better maintainability, with the strongest improvement observed in \textit{w/o-Testing} (–19\%). However, reliability consistently worsens across all variants, indicating that the generated code becomes slightly less stable.
In contrast, both \textbf{deepseek-chat} and \textbf{claude-3.5-haiku} display mixed results for maintainability: certain variants, such as \textit{w/o-Testing}, show moderate improvement, while others, like \textit{Full} or \textit{w/o-Design}, slightly regress. Despite these fluctuations, both models share a common trend of reduced reliability across all Waterfall settings, suggesting that the stage-by-stage process may help maintain code organization but at the expense of overall stability.

\uhead{Clean Code Attributes.} \textbf{\textit{Consistency increases across all models under Waterfall approaches while Intentionality shows mixed results. Adaptability clearly decreases for both \emph{gpt-4o-mini} and \emph{deepseek-chat} but shows mixed outcomes for \emph{claude-3-5-haiku}.}} gpt-4o-mini achieves strong reductions in consistency issues (–32\% to –43\%), with mixed intentionality results and severe adaptability regressions (+200–390\%). Deepseek-chat improves less consistently (consistency –16\% to –35\%, intentionality best in w/o-Testing at –43\%, adaptability increases only +13–95\%). By contrast, claude-3.5-haiku offers balanced improvements, especially in w/o-Testing (–6\% consistency, –9\% intentionality, –18\% adaptability). Smaller models may behave this way because they follow each Waterfall step too strictly, which boosts consistency but limits adaptability.

\begin{tcolorbox}
\textbf{RQ1 Summary:} Waterfall workflows improve software quality and code cleanliness but reduce adaptability, intentionality, and Pass@1 accuracy with more reliability issues. Their effectiveness also varies by model; \texttt{gpt-4o-mini} and \texttt{deepseek-chat} perform worse, whereas \texttt{claude-3.5-haiku} shows significant improvement.
\end{tcolorbox}

\FloatBarrier
\begin{table*}[t!]
\centering
\scriptsize
\caption{Comparison of error types across three different models. Each value reports the raw count and relative change (\%) compared to the model’s \textit{RawPrompt} baseline across Waterfall variants.}
\label{tab:runtime error frequency}
\renewcommand{\arraystretch}{1}
\setlength{\tabcolsep}{3pt}
\begin{tabular}{lccccccccccccccc}
\hline
\multirow{2}{*}{\textbf{Error Type}} &
\multicolumn{5}{c}{\textbf{gpt-4o-mini}} &
\multicolumn{5}{c}{\textbf{deepseek-chat}} &
\multicolumn{5}{c}{\textbf{claude-3.5-haiku}} \\ 
\cmidrule(lr){2-6} \cmidrule(lr){7-11} \cmidrule(lr){12-16}

& \textbf{Raw} & \multicolumn{4}{c}{\textbf{Waterfall}} & \textbf{Raw} & \multicolumn{4}{c}{\textbf{Waterfall}} & \textbf{Raw} & \multicolumn{4}{c}{\textbf{Waterfall}} \\
\cmidrule(lr){3-6} \cmidrule(lr){8-11} \cmidrule(lr){13-16}
& & \textbf{Full} & \textbf{w/o-Req.} & \textbf{w/o-Des.} & \textbf{w/o-Test.} & & \textbf{Full} & \textbf{w/o-Req.} & \textbf{w/o-Des.} & \textbf{w/o-Test.} & & \textbf{Full} & \textbf{w/o-Req.} & \textbf{w/o-Des.} & \textbf{w/o-Test.}\\

\hline
AssertionError & 53 & \cellcolor{red!13}{60 (+13\%)} & \cellcolor{green!3}{54 (+2\%)} & \cellcolor{red!17}{62 (+17\%)} & \cellcolor{green!6}{51 (-4\%)} &
                 51 & \cellcolor{green!2}{50 (-2\%)} & \cellcolor{red!8}{55 (+8\%)} & \cellcolor{red!14}{58 (+14\%)} & \cellcolor{red!18}{60 (+18\%)} &
                 27 & \cellcolor{red!50}{58 (+115\%)} & \cellcolor{red!50}{61 (+126\%)} & \cellcolor{red!50}{54 (+100\%)} & \cellcolor{red!50}{54 (+100\%)} \\
ValueError & 7 & \cellcolor{red!50}{25 (+257\%)} & \cellcolor{red!50}{24 (+243\%)} & \cellcolor{red!50}{22 (+214\%)} & \cellcolor{red!50}{28 (+300\%)} &
               7 & \cellcolor{red!50}{24 (+243\%)} & \cellcolor{red!50}{17 (+143\%)} & \cellcolor{red!50}{15 (+114\%)} & \cellcolor{red!50}{29 (+314\%)} &
               3 & \cellcolor{red!50}{22 (+633\%)} & \cellcolor{red!50}{17 (+467\%)} & \cellcolor{red!50}{17 (+467\%)} & \cellcolor{red!50}{25 (+733\%)} \\
TypeError & 4 & \cellcolor{red!29}{7 (+75\%)} & \cellcolor{red!38}{8 (+100\%)} & \cellcolor{red!44}{9 (+125\%)} & \cellcolor{red!44}{9 (+125\%)} &
             5 & \cellcolor{green!17}{4 (-20\%)} & \cellcolor{green!17}{4 (-20\%)} & 5 & \cellcolor{red!28}{7 (+40\%)} &
             1 & \cellcolor{red!33}{3 (+200\%)} & \cellcolor{red!38}{4 (+300\%)} & \cellcolor{red!44}{5 (+400\%)} & \cellcolor{red!33}{3 (+200\%)} \\
AttributeError & 5 & 5 & \cellcolor{red!10}{6 (+20\%)} & \cellcolor{red!10}{6 (+20\%)} & \cellcolor{red!50}{10 (+100\%)} &
                 3 & 3 & \cellcolor{green!17}{2 (-33\%)} & \cellcolor{red!17}{4 (+33\%)} & 3 &
                 1 & \cellcolor{red!33}{2 (+100\%)} & \cellcolor{red!33}{2 (+100\%)} & \cellcolor{red!50}{4 (+300\%)} & \cellcolor{red!33}{2 (+100\%)} \\
KeyError & 0 & \cellcolor{red!50}{4 (+400\%)} & \cellcolor{red!50}{4 (+400\%)} & \cellcolor{red!50}{2 (+200\%)} & \cellcolor{red!50}{3 (+300\%)} &
             1 & \cellcolor{red!50}{3 (+200\%)} & \cellcolor{red!33}{2 (+100\%)} & \cellcolor{red!33}{3 (+100\%)} & \cellcolor{red!33}{3 (+100\%)} &
             0 & \cellcolor{red!50}{2 (+200\%)} & \cellcolor{red!50}{4 (+400\%)} & \cellcolor{red!50}{5 (+500\%)} & \cellcolor{red!33}{1 (+100\%)} \\
FileNotFoundError & 2 & \cellcolor{red!38}{5 (+150\%)} & \cellcolor{red!19}{3 (+50\%)} & \cellcolor{red!19}{3 (+50\%)} & \cellcolor{red!25}{4 (+100\%)} &
                     1 & \cellcolor{red!33}{2 (+100\%)} & \cellcolor{red!50}{3 (+200\%)} & \cellcolor{red!33}{2 (+100\%)} & \cellcolor{red!50}{3 (+200\%)} &
                     1 & \cellcolor{red!33}{2 (+100\%)} & 1 & \cellcolor{red!33}{2 (+100\%)} & 1 \\
IndexError & 2 & \cellcolor{red!33}{4 (+100\%)} & \cellcolor{green!50}{0 (-100\%)} & \cellcolor{green!50}{0 (-100\%)} & \cellcolor{green!17}{1 (-50\%)} &
               2 & \cellcolor{green!50}{0 (-100\%)} & \cellcolor{green!50}{0 (-100\%)} & \cellcolor{green!50}{0 (-100\%)} & \cellcolor{green!17}{1 (-50\%)} &
               1 & 1 & \cellcolor{green!50}{0 (-100\%)} & 1 & \cellcolor{green!50}{0 (-100\%)} \\
NameError & 3 & \cellcolor{green!25}{2 (-33\%)} & \cellcolor{green!50}{1 (-67\%)} & \cellcolor{green!50}{1 (-67\%)} & 3 &
             2 & \cellcolor{green!33}{1 (-50\%)} & 2 & 2 & 2 &
             51 & \cellcolor{green!50}{3 (-94\%)} & \cellcolor{green!50}{2 (-96\%)} & \cellcolor{green!50}{2 (-96\%)} & \cellcolor{green!17}{20 (-61\%)} \\
SyntaxError & 0 & 0 & \cellcolor{red!50}{1 (+100\%)} & \cellcolor{red!50}{2 (+200\%)} & \cellcolor{red!50}{2 (+200\%)} &
               1 & \cellcolor{red!50}{5 (+400\%)} & \cellcolor{red!50}{6 (+500\%)} & \cellcolor{red!50}{5 (+400\%)} & \cellcolor{red!50}{6 (+500\%)} &
               0 & 0 & \cellcolor{red!50}{4 (+400\%)} & \cellcolor{red!50}{2 (+200\%)} & \cellcolor{red!50}{1 (+100\%)} \\
ZeroDivisionError & 2 & \cellcolor{green!50}{0 (-100\%)} & \cellcolor{green!50}{0 (-100\%)} & \cellcolor{green!50}{0 (-100\%)} & \cellcolor{green!50}{0 (-100\%)} &
                     3 & \cellcolor{green!50}{0 (-100\%)} & \cellcolor{green!50}{0 (-100\%)} & \cellcolor{green!50}{0 (-100\%)} & \cellcolor{green!50}{0 (-100\%)} &
                     2 & 2 & \cellcolor{green!17}{1 (-50\%)} & \cellcolor{green!17}{1 (-50\%)} & \cellcolor{green!50}{0 (-100\%)} \\
OperationalError* & 0 & \cellcolor{red!33}{1 (+100\%)} & \cellcolor{red!50}{2 (+200\%)} & \cellcolor{red!50}{2 (+200\%)} & \cellcolor{red!33}{1 (+100\%)} &
                           1 & \cellcolor{red!33}{2 (+100\%)} & \cellcolor{green!50}{0 (-100\%)} & \cellcolor{green!50}{0 (-100\%)} & \cellcolor{red!33}{1 (+100\%)} &
                           0 & \cellcolor{red!33}{1 (+100\%)} & \cellcolor{red!50}{2 (+200\%)} & \cellcolor{red!33}{1 (+100\%)} & \cellcolor{red!33}{1 (+100\%)} \\
Time Limit Exceeded & 2 & \cellcolor{green!25}{1 (-50\%)} & \cellcolor{green!25}{1 (-50\%)} & \cellcolor{green!50}{0 (-100\%)} & \cellcolor{green!50}{0 (-100\%)} &
                      2 & \cellcolor{green!50}{0 (-100\%)} & \cellcolor{green!25}{1 (-50\%)} & \cellcolor{green!25}{1 (-50\%)} & 2 &
                      1 & \cellcolor{green!50}{0 (-100\%)} & 1 & 1 & 1 \\
Exception & 0 & \cellcolor{red!33}{1 (+100\%)} & \cellcolor{red!33}{1 (+100\%)} & 0 & \cellcolor{red!33}{1 (+100\%)} &
              0 & 0 & 0 & 0 & \cellcolor{red!50}{2 (+200\%)} &
              0 & \cellcolor{red!33}{1 (+100\%)} & 0 & 0 & 0 \\
DeprecationError* & 1 & \cellcolor{green!50}{0 (-100\%)} & \cellcolor{green!50}{0 (-100\%)} & \cellcolor{green!50}{0 (-100\%)} & 1 &
                                 1 & 1 & 1 & \cellcolor{green!50}{0 (-100\%)} & 1 &
                                 0 & 0 & 1 & 1 & 0 \\
OSError & 0 & \cellcolor{red!33}{1 (+100\%)} & 0 & 0 & 0 &
           1 & \cellcolor{green!50}{0 (-100\%)} & 0 & 0 & 1 &
           1 & \cellcolor{green!50}{0 (-100\%)} & 1 & 1 & 1 \\
One-Off Errors & 1 & \cellcolor{red!50}{6 (+500\%)} & 1 & \cellcolor{green!50}{0 (-100\%)} & \cellcolor{red!33}{4 (+300\%)} &
                 0 & \cellcolor{red!33}{1 (+100\%)} & \cellcolor{red!33}{1 (+100\%)} & 0 & \cellcolor{red!50}{3 (+300\%)} &
                 0 & \cellcolor{red!33}{2 (+200\%)} & \cellcolor{red!33}{1 (+100\%)} & \cellcolor{red!33}{1 (+100\%)} & \cellcolor{red!33}{1 (+100\%)} \\
\hline
Total           & 82  & \cellcolor{red!49}{122 {\scriptsize(+49\%)}} & \cellcolor{red!29}{106 {\scriptsize(+29\%)}} & \cellcolor{red!33}{109 {\scriptsize(+33\%)}} & \cellcolor{red!44}{118 {\scriptsize(+44\%)}} 
                & 81  & \cellcolor{red!19}{96 {\scriptsize(+19\%)}} & \cellcolor{red!16}{94 {\scriptsize(+16\%)}} & \cellcolor{red!17}{95 {\scriptsize(+17\%)}} & \cellcolor{red!53}{124 {\scriptsize(+53\%)}} 
                & 89  & \cellcolor{red!11}{99 {\scriptsize(+11\%)}} & \cellcolor{red!15}{102 {\scriptsize(+15\%)}} & \cellcolor{red!10}{98 {\scriptsize(+10\%)}} & \cellcolor{red!25}{111 {\scriptsize(+25\%)}}\\
\hline
\end{tabular}
\vspace{2pt}

\textit{\textbf{Note:}} \textit{PyPDF2.errors.DeprecationError} is shortened to DeprecationError*. \textit{sqlite3.OperationalError} is shortened to OperationalError*.

\end{table*}

\subsection{RQ2: What kinds of errors occur during class-level code generation?}

\phead{Motivation:}
In this research question, we focus on identifying and understanding the types of errors that arise during class-level code generation when following the Waterfall process model. In RQ1, we found that although Waterfall-based workflows improved code quality and cleanliness, they also led to more runtime reliability issues and stage inconsistencies. Building on these findings, here we investigate why such failures occur by examining the types of runtime errors that emerge during class-level code generation. By categorizing and comparing these errors across different LLMs and workflow variants, we aim to uncover the underlying causes of coordination breakdowns.

\phead{Approach:}
We manually evaluated all 1,500 generated code samples (100 ClassEval tasks × 3 models × 5 workflows) to identify the most frequent error types. Each code sample was executed against its corresponding unit tests from the ClassEval benchmark, and the resulting errors were recorded. Error types were categorized following methodologies from prior studies\cite{lin2024soen,qian2023chatdev}. Specifically, SOEN-101 \cite{lin2024soen} classified failure types according to the official Python Interpreter documentation\footnote{\url{https://docs.python.org/3/library/exceptions.html}}, while Chatdev \cite{qian2023chatdev} identified successful compilations and recorded compiler-reported errors based on feedback logs between the testing process.
During testing, simple syntax errors, such as missing or extra parentheses and full stops, were corrected manually by reviewing the code, and the evaluation was rerun. Repeated occurrences of the same error type within a single class (e.g., multiple \textit{AssertionError}s) were counted once to prevent overcounting. Two extra categories were added: (a) \textit{Time Limit Exceeded}, for tests running longer than eight minutes, and (b) \textit{One-Off Errors}, grouping rare errors that appeared only once or twice per model across all workflows, like \texttt{ModuleNotFoundError}, \texttt{re.error}, \texttt{RuntimeError} etc.
Finally, we compiled the results into an error frequency chart (Table~\ref{tab:runtime error frequency}), where each frequency value represents the number of classes in which that error occurred.

\phead{Results:}
\textbf{\textit{Among the evaluated workflows, three error types \textit{ValueError}, \textit{AssertionError}, and \textit{TypeError} account for most failures, each showing substantial growth when using Waterfall (ranging from about +40\% to over +700\%).}}
\textit{ValueError} increases the most among all error types, increasing from around 7 to 25 cases in several configurations
\textit{AssertionError} also increases, in some cases by more than 100\%. \textit{TypeError} typically doubles.
Meanwhile, simpler runtime issues decline sharply. \textit{ZeroDivisionError} nearly disappears (reductions of up to –100\%), \textit{IndexError} often drops to zero, and \textit{NameError} decreases by over 90\% in several workflows.
\textit{KeyError} increases from almost none to as many as five occurrences per workflow (about +100–500\%), and both \textit{OperationalError} and grouped \textit{One-Off Errors} also rise. 

\textbf{\textit{Across all models, introducing the Waterfall workflow increases total runtime errors by 10–53\% compared to the RawPrompt baseline.}}
The overall rise in errors is most pronounced for \emph{GPT-4o-mini} (+49\%), \emph{DeepSeek-Chat} displays a more balanced response (+16–19\%). In contrast, \emph{Claude-3.5-haiku} shows a limited overall increase (+10–25\%).
Notably, removing the testing stage (\textit{w/o-Test}) consistently leads to the highest failure rates across all models up to +53\% and shows the importance of test awareness. 

\textbf{\textit{Across workflow variants, the full Waterfall process leads to the highest number of runtime errors, while removing stages gives mixed and model-specific effects.}}
In most cases, the full Waterfall setup raises failures compared to Raw Prompt. For example, \emph{GPT-4o-mini} has about 40 more errors under Waterfall (Full) (82 to 122, +49\%), and \emph{DeepSeek-Chat} adds nearly 15 (+19\%) compared to when using Raw Prompt.
When stages are removed, changes are small and inconsistent. Skipping \textit{requirements} (w/o-Req.) or \textit{design} (w/o-Des.) gives slight drops for some models. \emph{DeepSeek-Chat}, for instance, falls by only a few errors, while \emph{Claude-3.5-haiku} is almost unchanged.
The clearest pattern is with \textit{w/o-Test}. It yields the biggest increase across models, reaching 124 errors for \emph{DeepSeek-Chat} and 118 for \emph{GPT-4o-mini}. Across all workflows,\textit{w/o-Test} performs the worst, \textit{w/o-Req.} and \textit{w/o-Des.} stay within 5\% difference of each other but \textit{Full} varies wildly compared to baseline.
Overall, each stage affects reliability differently. Removing early stages has little impact, but skipping testing consistently makes the system much more error-prone.

\begin{tcolorbox}
\textbf{RQ2 Summary:} Waterfall workflows increase runtime errors by 10–53\% compared to RawPrompt. \textit{ValueError}, \textit{AssertionError}, and \textit{TypeError} dominate the failures, while \textit{ZeroDivisionError} and other low-level mistakes decline. \texttt{DeepSeek-Chat} is the comparatively more stable model, whereas \texttt{GPT-4o-mini} and \texttt{Claude-3.5-haiku} show greater sensitivity to workflow design, indicating that structured prompting must be carefully tuned for each model.
\end{tcolorbox}

\subsection{RQ3: What are the underlying causes behind these failures?}

\phead{Motivation:}  
While RQ2 identified the types of runtime errors produced during multi-agent code generation, it did not explain why these errors occurred. In this research question, we aim to identify the primary reasons behind these failures by examining how the various stages of the workflow interact with each other. Specifically, we examine whether the errors arise from missing or inconsistent information passed between agents, incorrect implementation of design decisions, or limitations in the models’ reasoning abilities. By identifying these causes, we can gain a deeper understanding of how workflow structure, model behaviour, and stage coordination contribute to the overall reliability of multi-agent code generation.

\begin{table}[t]
\centering
\scriptsize
\caption{Taxonomy of error categories with corresponding motivations describing the underlying causes of failures in class-level code generation.}
\label{tab:error taxonomy definition}
\renewcommand{\arraystretch}{1}
\begin{tabular}{p{2.8cm} p{5cm}}
\toprule
\textbf{Error Category} & \textbf{Motivation} \\
\midrule

\textbf{Missing Code} & Code components such as variables, functions, or classes are missing, renamed, or only partially implemented. \\
\hspace*{1em}Renamed Variable & A variable name has been changed from what the specification requires. \\
\hspace*{1em}Missing Variable & A required variable from the specification is missing in the code. \\
\hspace*{1em}Missing Function & A function is called but not defined anywhere in the code. \\
\hspace*{1em}Renamed Function & A function was renamed, but the test still uses the old name. \\
\hspace*{1em}Missing Class & A class used in the code/test is not defined. \\
\hspace*{1em}Renamed Class & A class was renamed but tests still use the old name. \\
\hspace*{1em}Missing Import & Code uses a module or function without importing it. \\
\hspace*{1em}Incomplete Implementation & Only part of the instructed functionality is implemented. \\

\midrule

\textbf{Return Mismatch} & Returned outputs differ from expected results due to incorrect arity, order, type, or format. \\
\hspace*{1em}Arity Mismatch & Function returns too many or too few values (e.g., 3 items instead of 2). \\
\hspace*{1em}Order Mismatch & Function returns values in the wrong order (e.g., (b, a) instead of (a, b)). \\
\hspace*{1em}Type Mismatch & The return value is of the wrong Python type (e.g., list instead of tuple). \\
\hspace*{1em}Format Mismatch & The return value has incorrect formatting (e.g., casing, spacing, symbols). \\

\midrule

\textbf{Input Validation} & The code fails to correctly handle user input due to being too strict, too lenient, or logically flawed. \\
\hspace*{1em}Overrestrictive Validation & Adds unnecessary constraints not required by the problem. \\
\hspace*{1em}Faulty Validation & Validation logic is incorrect or incomplete. \\
\hspace*{1em}Missing Input Validation & No validation occurs where it should. \\

\midrule

\textbf{Semantic Failure} & Logical or algorithmic errors causing functionality to deviate from intended behavior. \\
\hspace*{1em}Spec Violation & Code violates explicit problem instructions. \\
\hspace*{1em}Signature Mismatch & Function parameters differ from the expected specification. \\
\hspace*{1em}Wrong Algorithm & Code implements an incorrect algorithm or formula. \\
\hspace*{1em}Wrong Edge Case Handling & Fails to handle boundary cases properly. \\
\hspace*{1em}Timeout & Code runs indefinitely or exceeds time limits. \\
\hspace*{1em}Syntax Error & Code cannot compile or parse due to syntax issues. \\
\hspace*{1em}Integration Error & A newly added feature conflicts with existing code. \\

\midrule

\textbf{Dataset} & Errors arise from ambiguous, flawed, or inconsistent problem specifications or test data. \\
\hspace*{1em}Faulty Spec & The problem description is incorrect or misleading. \\
\hspace*{1em}Faulty Test & Test cases contain incorrect or inconsistent logic. \\
\hspace*{1em}Missing Import (Test) & Test scripts fail due to missing imports. \\
\hspace*{1em}Spec-Test Mismatch & Tests contradict the written problem specification. \\

\midrule

\textbf{Environment} & Failures caused by dependency, compatibility, or versioning issues. \\
\hspace*{1em}Version Incompatibility & Uses unsupported features in the given Python version. \\
\hspace*{1em}Undeclared Dependency & Uses undeclared external packages. \\
\hspace*{1em}Deprecated Dependency & Uses outdated or removed dependencies. \\

\bottomrule
\end{tabular}
\end{table}

\phead{Approach:}
We first isolated all failing classes from the 1,500 code samples analyzed in RQ2 (100 ClassEval tasks × 3 models × 5 workflows). Each failed class was manually inspected to identify the root cause of the failure. For every sample, we compared the generated code against three main references:

\uheadu{ClassEval Task Description.}
We examined the problem description in ClassEval to check whether the generated code met the specified requirements and implemented the intended functionality. We also reviewed the task descriptions to identify any ambiguities, inconsistencies, or missing details that could affect code generation.

\uheadu{Unit Tests.}
We reviewed the corresponding unit tests to identify which behaviors failed during execution and to determine whether the issue came from missing functionality, logic errors, or incorrect handling of test cases. We also examined whether the unit tests were consistent with the problem description and verified that they did not contain contradictions or inherent design flaws.

\uheadu{Reference Implementation.}
We compared the generated code with the canonical reference implementation to detect deviations in logic, data flow, and structural design that led to incorrect behavior. We also examined the reference implementation to determine whether it provided additional details beyond those stated in the problem description.

Following prior work on LLM fault analysis \cite{xue2025classeval, qian2023chatdev}, we labeled each identified issue using a structured taxonomy of error categories. These categories were grouped into broader types, as summarized in Table~\ref{tab:error taxonomy definition}. Each category includes fine-grained subtypes and a brief explanation of the issue's nature. Each class was assigned one primary cause to avoid duplication, but when a failure stemmed from several interrelated problems, it was cross-referenced with multiple relevant categories. We applied this process across \texttt{gpt-4o-mini}, \texttt{deepseek-chat}, and \texttt{claude-3.5-haiku} under the agent-based Waterfall workflows.

The taxonomy covers a wide spectrum of errors commonly observed in LLM-generated code \cite{yuenprompting}. \textit{Missing Code} and \textit{Return Mismatch} represent structural gaps in the code (e.g., missing or renamed functions, variables, classes, etc.) and output inconsistencies. \textit{Input Validation} and \textit{Semantic Failure} capture reasoning and logical flaws, such as invalid checks or incorrect algorithms. \textit{Dataset} and \textit{Environment} categories describe issues related to problem and test design, dependencies, or version incompatibilities.

\phead{Results:} 
\begin{table*}[t]
\centering
\scriptsize
\caption{Qualitative analysis of error categories across models and Waterfall variants. Each cell shows the relative frequency of errors compared to the \textit{RawPrompt} baseline, where red indicates an increase and green indicates a decrease.}
\label{tab:error taxonomy detail}
\renewcommand{\arraystretch}{0.85}
\begin{tabular}{p{4cm} >{\centering\arraybackslash}p{0.5cm} >{\centering\arraybackslash}p{0.5cm} >{\centering\arraybackslash}p{0.5cm} >{\centering\arraybackslash}p{0.5cm} >{\centering\arraybackslash}p{0.5cm} >{\centering\arraybackslash}p{0.5cm} >{\centering\arraybackslash}p{0.5cm} >{\centering\arraybackslash}p{0.5cm} >{\centering\arraybackslash}p{0.5cm} >{\centering\arraybackslash}p{0.5cm} >{\centering\arraybackslash}p{0.5cm} >{\centering\arraybackslash}p{0.5cm} >{\centering\arraybackslash}p{0.5cm} >{\centering\arraybackslash}p{0.5cm} >{\centering\arraybackslash}p{0.5cm}}
\toprule
\textbf{Error Category} & \multicolumn{5}{c}{\textbf{gpt-4o-mini}} & \multicolumn{5}{c}{\textbf{deepseek-chat}} & \multicolumn{5}{c}{\textbf{claude-3-5-haiku}} \\
\cmidrule(lr){2-6} \cmidrule(lr){7-11} \cmidrule(lr){12-16}
& \textbf{Raw} & \multicolumn{4}{c}{\textbf{Waterfall}} & \textbf{Raw} & \multicolumn{4}{c}{\textbf{Waterfall}} & \textbf{Raw} & \multicolumn{4}{c}{\textbf{Waterfall}} \\
\cmidrule(lr){3-6} \cmidrule(lr){8-11} \cmidrule(lr){13-16}
& & \textbf{Full} & \textbf{w/o-Req.} & \textbf{w/o-Des.} & \textbf{w/o-Test.} & & \textbf{Full} & \textbf{w/o-Req.} & \textbf{w/o-Des.} & \textbf{w/o-Test.} & & \textbf{Full} & \textbf{w/o-Req.} & \textbf{w/o-Des.} & \textbf{w/o-Test.}\\
\midrule


\multicolumn{1}{l}{\textbf{Missing Code}} & \textbf{4} & \improve{\textbf{1}}{75} & \neutral{\textbf{4}} & \improve{\textbf{0}}{100} & \regress{\textbf{10}}{100} 
& \textbf{1} & \improve{\textbf{0}}{100} & \improve{\textbf{0}}{100} & \neutral{\textbf{1}} & \neutral{\textbf{1}}
& \textbf{50} & \improve{\textbf{2}}{100} & \improve{\textbf{1}}{100} & \improve{\textbf{1}}{100} & \improve{\textbf{17}}{75}\\
\hspace*{1em}Renamed Variable & 1 & \neutral{1} & \regress{2}{50} & \improve{0}{100} & \improve{0}{100}
& 0 & \neutral{0} & \neutral{0} & \neutral{0} & \neutral{0}
& 0 & \neutral{0} & \neutral{0} & \neutral{0} & \neutral{0}\\
\hspace*{1em}Missing Variable & 0 & \neutral{0} & \regress{1}{100} & \neutral{0} & \neutral{0}
& 0 & \neutral{0} & \neutral{0} & \regress{1}{100} & \neutral{0}
& 0 & \neutral{0} & \neutral{0} & \neutral{0} & \neutral{0}\\
\hspace*{1em}Missing Function & 0 & \neutral{0} & \neutral{0} & \neutral{0} & \regress{2}{100}
& 0 & \neutral{0} & \neutral{0} & \neutral{0} & \neutral{0}
& 0 & \neutral{0} & \neutral{0} & \neutral{0} & \neutral{0}\\
\hspace*{1em}Renamed Function & 0 & \neutral{0} & \neutral{0} & \neutral{0} & \regress{6}{100}
& 1 & \improve{0}{100} & \improve{0}{100} & \improve{0}{100} & \neutral{1}
& 0 & \neutral{0} & \neutral{0} & \neutral{0} & \neutral{0}\\
\hspace*{1em}Missing Class & 1 & \improve{0}{100} & \improve{0}{100} & \improve{0}{100} & \improve{0}{100}
& 0 & \neutral{0} & \neutral{0} & \neutral{0} & \neutral{0}
& 49 & \improve{2}{100} & \improve{1}{100} & \improve{1}{100} & \improve{17}{75}\\
\hspace*{1em}Renamed Class & 0 & \neutral{0} & \neutral{0} & \neutral{0} & \regress{1}{100}
& 0 & \neutral{0} & \neutral{0} & \neutral{0} & \neutral{0}
& 0 & \neutral{0} & \neutral{0} & \neutral{0} & \neutral{0}\\
\hspace*{1em}Missing Import & 2 & \improve{0}{100} & \improve{0}{100} & \improve{0}{100} & \improve{0}{100}
& 0 & \neutral{0} & \neutral{0} & \neutral{0} & \neutral{0}
& 1 & \improve{0}{100} & \improve{0}{100} & \improve{0}{100} & \improve{0}{100}\\
\hspace*{1em}Incomplete Implementation & 0 & \neutral{0} & \regress{1}{100} & \neutral{0} & \regress{1}{100}
& 0 & \neutral{0} & \neutral{0} & \neutral{0} & \neutral{0}
& 0 & \neutral{0} & \neutral{0} & \neutral{0} & \neutral{0}\\

\addlinespace[1em]

\multicolumn{1}{l}{\textbf{Return Mismatch}} & \textbf{23} & \regress{\textbf{26}}{13} & \regress{\textbf{28}}{22} & \regress{\textbf{27}}{17} & \regress{\textbf{26}}{13}
& \textbf{21} & \regress{\textbf{23}}{10} & \regress{\textbf{25}}{19} & \regress{\textbf{34}}{61} & \regress{\textbf{32}}{52}
& \textbf{13} & \regress{\textbf{22}}{69} & \regress{\textbf{27}}{108} & \regress{\textbf{27}}{108} & \regress{\textbf{26}}{100}\\
\hspace*{1em}Arity Mismatch & 4 & \improve{1}{75} & \regress{6}{50} & \improve{3}{25} & \improve{3}{25}
& 5 & \improve{4}{20} & \neutral{5} & \regress{7}{40} & \improve{3}{40}
& 3 & \regress{5}{67} & \regress{5}{67} & \improve{4}{33} & \improve{4}{33}\\
\hspace*{1em}Order Mismatch & 1 & \regress{2}{50} & \neutral{1} & \improve{0}{100} & \neutral{1}
& 1 & \improve{0}{100} & \neutral{1} & \neutral{1} & \neutral{1}
& 1 & \improve{0}{100} & \regress{3}{200} & \regress{3}{200} & \neutral{1}\\
\hspace*{1em}Type Mismatch & 8 & \regress{13}{63} & \regress{14}{75} & \regress{13}{63} & \regress{16}{100}
& 8 & \improve{7}{13} & \regress{9}{13} & \regress{11}{38} & \regress{13}{63}
& 3 & \regress{5}{67} & \regress{7}{133} & \regress{9}{200} & \regress{7}{133}\\
\hspace*{1em}Format Mismatch & 10 & \neutral{10} & \improve{7}{30} & \regress{11}{10} & \improve{6}{40}
& 7 & \regress{12}{71} & \regress{10}{43} & \regress{15}{114} & \regress{14}{100}
& 6 & \regress{12}{100} & \regress{12}{100} & \regress{11}{83} & \regress{14}{133}\\

\addlinespace[1em]

\multicolumn{1}{l}{\textbf{Input Validation}} & \textbf{2} & \regress{\textbf{14}}{600} & \regress{\textbf{16}}{700} & \regress{\textbf{10}}{400} & \regress{\textbf{13}}{550}
& \textbf{3} & \regress{\textbf{9}}{200} & \regress{\textbf{8}}{167} & \regress{\textbf{8}}{167} & \regress{\textbf{18}}{500}
& \textbf{1} & \regress{\textbf{10}}{900} & \regress{\textbf{6}}{500} & \regress{\textbf{3}}{200} & \regress{\textbf{11}}{1000}\\
\hspace*{1em}Overrestrictive Validation & 0 & \regress{9}{100} & \regress{10}{100} & \regress{6}{100} & \regress{7}{100}
& 0 & \regress{6}{100} & \regress{5}{100} & \regress{4}{100} & \regress{12}{100}
& 0 & \regress{6}{100} & \regress{3}{100} & \regress{2}{100} & \regress{7}{100}\\
\hspace*{1em}Faulty Validation & 2 & \regress{5}{150} & \regress{6}{200} & \regress{4}{100} & \regress{6}{200}
& 1 & \regress{3}{200} & \regress{2}{100} & \regress{2}{100} & \regress{6}{500}
& 1 & \regress{3}{200} & \regress{3}{200} & \neutral{1} & \regress{4}{300}\\
\hspace*{1em}Missing Input Validation & 0 & \neutral{0} & \neutral{0} & \neutral{0} & \neutral{0}
& 2 & \improve{0}{100} & \improve{1}{50} & \improve{2}{0} & \improve{0}{100}
& 0 & \regress{1}{100} & \neutral{0} & \neutral{0} & \neutral{0}\\

\addlinespace[1em]

\multicolumn{1}{l}{\textbf{Semantic Failure}} & \textbf{35} & \regress{\textbf{51}}{46} & \regress{\textbf{49}}{40} & \regress{\textbf{55}}{57} & \regress{\textbf{50}}{43}
& \textbf{29} & \regress{\textbf{52}}{79} & \regress{\textbf{50}}{72} & \regress{\textbf{47}}{62} & \regress{\textbf{53}}{83}
& \textbf{18} & \regress{\textbf{43}}{139} & \regress{\textbf{54}}{200} & \regress{\textbf{51}}{183} & \regress{\textbf{39}}{117}\\
\hspace*{1em}Spec Violation & 2 & \regress{12}{500} & \regress{7}{250} & \regress{9}{350} & \regress{7}{250}
& 1 & \regress{5}{400} & \regress{3}{200} & \regress{3}{200} & \regress{9}{800}
& 0 & \regress{4}{100} & \regress{4}{100} & \regress{3}{100} & \regress{2}{100}\\
\hspace*{1em}Signature Mismatch & 0 & \neutral{0} & \regress{1}{100} & \regress{2}{100} & \regress{3}{100}
& 0 & \neutral{0} & \neutral{0} & \neutral{0} & \regress{1}{100}
& 0 & \neutral{0} & \neutral{0} & \regress{1}{100} & \neutral{0}\\
\hspace*{1em}Wrong Algorithm & 22 & \regress{25}{14} & \regress{25}{14} & \regress{25}{14} & \improve{23}{5}
& 19 & \regress{24}{26} & \regress{22}{16} & \regress{24}{26} & \regress{23}{21}
& 13 & \regress{29}{123} & \regress{26}{100} & \regress{30}{131} & \regress{23}{77}\\
\hspace*{1em}Wrong Edge Case Handling & 9 & \regress{10}{11} & \improve{8}{11} & \regress{12}{33} & \improve{8}{11}
& 4 & \regress{5}{25} & \regress{5}{25} & \regress{5}{25} & \improve{1}{75}
& 4 & \regress{6}{50} & \regress{9}{125} & \regress{7}{75} & \regress{7}{75}\\
\hspace*{1em}Timeout & 2 & \improve{1}{50} & \improve{1}{50} & \improve{0}{100} & \improve{0}{100}
& 2 & \improve{0}{100} & \improve{1}{50} & \improve{0}{100} & \neutral{2}
& 1 & \improve{0}{100} & \neutral{1} & \neutral{1} & \neutral{1}\\
\hspace*{1em}Syntax Error & 0 & \neutral{0} & \regress{3}{100} & \regress{6}{100} & \regress{6}{100}
& 3 & \regress{14}{367} & \regress{17}{467} & \regress{13}{333} & \regress{14}{367}
& 0 & \regress{1}{100} & \regress{12}{100} & \regress{6}{100} & \regress{3}{100}\\
\hspace*{1em}Integration Error & 0 & \regress{3}{100} & \regress{4}{100} & \regress{1}{100} & \regress{3}{100}
& 0 & \regress{4}{100} & \regress{2}{100} & \regress{1}{100} & \regress{3}{100}
& 0 & \regress{3}{100} & \regress{2}{100} & \regress{3}{100} & \regress{3}{100}\\

\addlinespace[1em]

\multicolumn{1}{l}{\textbf{Dataset}} & \textbf{47} & \regress{\textbf{59}}{26} & \regress{\textbf{51}}{9} & \regress{\textbf{62}}{32} & \regress{\textbf{52}}{11}
& \textbf{52} & \improve{\textbf{49}}{6} & \neutral{\textbf{52}} & \improve{\textbf{48}}{8} & \neutral{\textbf{52}}
& \textbf{24} & \regress{\textbf{49}}{104} & \regress{\textbf{59}}{146} & \regress{\textbf{51}}{112} & \regress{\textbf{44}}{83}\\
\hspace*{1em}Faulty Spec & 15 & \regress{22}{47} & \regress{20}{33} & \regress{26}{73} & \regress{21}{40}
& 19 & \neutral{19} & \regress{21}{11} & \regress{21}{11} & \regress{21}{11}
& 6 & \regress{22}{267} & \regress{24}{300} & \regress{22}{267} & \regress{20}{233}\\
\hspace*{1em}Faulty Test & 6 & \regress{7}{17} & \regress{7}{17} & \regress{7}{17} & \neutral{6}
& 9 & \improve{8}{11} & \neutral{9} & \improve{7}{22} & \neutral{9}
& 5 & \regress{8}{60} & \regress{10}{100} & \regress{7}{40} & \neutral{5}\\
\hspace*{1em}Missing Import (Test) & 0 & \regress{1}{100} & \neutral{0} & \neutral{0} & \regress{1}{100}
& 1 & \improve{0}{100} & \neutral{1} & \improve{0}{100} & \improve{0}{100}
& 0 & \neutral{0} & \neutral{0} & \neutral{0} & \neutral{0}\\

\addlinespace[0.5em]

\multicolumn{1}{l}{\textbf{Environment}} & \textbf{1} & \neutral{\textbf{1}} & \regress{\textbf{2}}{80} & \neutral{\textbf{1}} & \regress{\textbf{5}}{80} & \textbf{1} & \neutral{\textbf{1}} & \regress{\textbf{2}}{80} & \neutral{\textbf{1}} & \regress{\textbf{2}}{80} & \textbf{1} & \improve{\textbf{0}}{80} & \neutral{\textbf{1}} & \neutral{\textbf{1}} & \regress{\textbf{3}}{80} \\
\hspace*{1em}Version Incompatibility  & \neutral{0} & \neutral{0} & \regress{2}{80} & \regress{1}{80} & \regress{3}{80} & \neutral{0} & \neutral{0} & \neutral{0} & \neutral{0} & \regress{1}{80} & \neutral{0} & \neutral{0} & \neutral{0} & \neutral{0} & \neutral{0} \\
\hspace*{1em}Undeclared Dependency    & \neutral{0} & \regress{1}{80} & \neutral{0} & \neutral{0} & \regress{1}{80} & \neutral{0} & \neutral{0} & \regress{1}{80} & \neutral{0} & \neutral{0} & 1 & \improve{0}{80} & \improve{0}{80} & \improve{0}{80} & \regress{3}{80} \\
\hspace*{1em}Deprecated Dependency    &1 & \improve{0}{80} & \improve{0}{80} & \improve{0}{80} & \neutral{1} & \neutral{1} & \neutral{1} & \neutral{1} & \improve{0}{80} & \neutral{1} & \neutral{0} & \neutral{0} & \regress{1}{80} & \regress{1}{80} & \neutral{0} \\

\bottomrule
\end{tabular}
\end{table*}
\textbf{\textit{Overall, \textit{Dataset}-related issues were the most common source of failure (47–62 cases), followed by reasoning and logic errors such as \textit{Semantic Failure} (50–55 cases) and output inconsistencies under \textit{Return Mismatch} (21–34 cases), with \textit{Input Validation} errors also increasing notably under Waterfall workflows.}}
As shown in Table~\ref{tab:error taxonomy detail}, these three categories \textit{Semantic Failure}, \textit{Return Mismatch}, and \textit{Dataset} consistently dominated across all Waterfall variants and models. This pattern reflects persistent challenges in algorithmic reasoning, output consistency, and handling of incomplete or ambiguous task inputs. In contrast, \textit{Input Validation} errors were rare in the RawPrompt baseline (2 cases) but became considerably more frequent under Waterfall workflows, rising to 18 cases. This increase suggests that while structured multi-agent workflows improve validation coverage, they can also surface deeper logical or constraint-related issues during testing, likely due to tighter coordination and stricter specification enforcement between stages.

\textbf{\textit{Across all models, a large number of the failures stemmed from reasoning and output errors. \textit{Semantic Failure} rose to 50–55 cases under Waterfall (from 35–49 in RawPrompt), while \textit{Return Mismatch} remained frequent (21–34 cases), mainly due to \textit{Type} and \textit{Format} issues ($\approx$75–83\% of return errors).}}
\textit{Semantic Failure} was one of the most common sources of errors, increasing sharply across all models: for \texttt{claude-3.5-haiku} from 18 to 54 cases (+36; $\approx$200\%), for \texttt{deepseek-chat} from 29 to 53 (+24; $\approx$83\%), and for \texttt{gpt-4o-mini} from 35 to 55 (+20; $\approx$57\%). 
Most stemmed from \textit{Wrong Algorithm} (19–30 cases) and \textit{Wrong Edge Case Handling} (4–12), indicating persistent weaknesses in logical precision and boundary reasoning; notably, \texttt{gpt-4o-mini} also showed a sharp increase in \textit{Specification Violations} (2→12). \textit{Return Mismatch} was also a common source of errors. For \texttt{gpt-4o-mini}, the number of these cases increased slightly by 3–5 (from 23 to 26–28, about 13–22\%). For \texttt{deepseek-chat}, the counts varied across Waterfall variants, ranging from 21 to 34. \texttt{Claude-3.5-haiku} showed the largest increase, rising errors by around 108\%. While \textit{Arity Mismatch} declined slightly in the full Waterfall setup, suggesting improved functional completeness, the coordination across stages also introduced new inconsistencies in return types and formatting, revealing a trade-off between structural rigor and semantic reliability.

\textbf{\textit{\textit{Input Validation} errors rose sharply under different Waterfall varaints, while \textit{Dataset} errors stayed high or increased depending on the model.}}
For \textit{Input Validation}, all models showed a notable rise under Waterfall. When using \texttt{gpt-4o-mini} as the LLM model, the error numbers increased from 2 to 16 cases, with the highest count observed when the \textit{requirements stage} was removed. \texttt{deepseek-chat} expanded from 3 to 18 ($\approx$500\%), peaking when the \textit{testing phase} was excluded, while using \texttt{claude-3.5-haiku}, the errors rose from a single case to 11 ($\approx$900\%), reaching its maximum in the full Waterfall setup. Most of these errors were \textit{Overrestrictive Validation}, such as unnecessary input checks added by the agents (\texttt{gpt-4o-mini}: 0–9; \texttt{deepseek-chat}: up to 12), or \textit{Faulty Validation}, where the conditions were incorrect or incomplete (\texttt{gpt-4o-mini}: 2–6; \texttt{deepseek-chat}: 1–6). \textit{Missing Validation} errors were rare, appearing only occasionally in \texttt{deepseek-chat} (0–2 cases) and \texttt{claude-3.5-haiku} (0–1), and entirely absent in \texttt{gpt-4o-mini}.
For \textit{Dataset} errors, \texttt{gpt-4o-mini} showed a moderate increase from 47 to 62 cases ($\approx$32\%), with the highest count appearing when the design phase was omitted. \texttt{deepseek-chat} remained relatively stable, fluctuating slightly between 49 and 52 cases across configurations. In contrast, \texttt{claude-3.5-haiku} showed a sharp rise from 24 in the baseline to 59 under Waterfall ($\approx$146\%), driven largely by \textit{Faulty Specifications} (rising from 6 to about 24) and, to a lesser extent, \textit{Faulty Tests} (5–10).

\textbf{\textit{\textit{Missing Code} errors dropped sharply under Waterfall workflows, falling from 4–50 cases in RawPrompt to as few as 1–2 across different models, while \textit{Environment} related issues remained low overall, typically between 1–5 cases.}}
For \textit{Missing Code}, \texttt{gpt-4o-mini} reduced the errors from 4 to 1, and \texttt{deepseek-chat} showed an even larger improvement, decreasing from 10 to just a single case compared to the raw prompt. The most pronounced reduction occurred in \texttt{claude-3.5-haiku}, where errors fell from 50 in the baseline to only 1–2 across most Waterfall variants, an improvement of over 95\%. These results indicate that the multi-stage structure of Waterfall workflows helped prevent incomplete implementations and improved overall code completeness.
In contrast, \textit{Environment} related issues showed minor fluctuations but stayed consistently low. \texttt{Claude-3.5-haiku} fully eliminated these problems in the full Waterfall configuration, while \texttt{gpt-4o-mini} and \texttt{deepseek-chat} showed small variations depending on which stage was removed. Subcategories like \textit{Version Incompatibility} and \textit{Undeclared Dependency} appeared only a few times, indicating that all models generally managed their dependencies and environments correctly.

\begin{tcolorbox}
\textbf{RQ3 Summary:}
Waterfall workflows reduce structural errors such as \textit{Missing Code} but increase reasoning and validation issues such as \textit{Semantic Failure} and \textit{Return Mismatch}. \textit{Dataset} related errors remain the most common overall. Each model show trade-offs; \texttt{gpt-4o-mini} and \texttt{deepseek-chat} improve completeness, while \texttt{claude-3.5-haiku} face higher logic and validation failures.
\end{tcolorbox}
\section{Discussion}

\subsection{Implications}
\phead{For Developers.} Waterfall-style, role-specialized agents reliably surface inconsistencies and promote disciplined design and validation practices, resulting in cleaner, more maintainable class implementations. However, certain Waterfall activities, particularly design and validation, introduce additional architectural structure and constraint checks, which improve code robustness and maintainability but simultaneously worsen execution success and functional accuracy by adding complexity and stricter logic not required by the original specification. The trade-off is that additional structure often introduces \emph{extra validation logic}, for example, tighter input checks, refactored interfaces, or modified boundary conditions that were not defined in the original requirements. These changes increase code robustness but can degrade functional accuracy (Pass@1) because benchmark test suites expect strict \emph{requirements conformance} rather than defensive or production-grade behaviour.

\emph{Practical guidance.} (i) \textit{Requirements-driven validation}: restrict defensive checks to what is explicitly stated in the task specification; treat any extended policy as optional. (ii) \textit{Stage optimization}: retain requirement clarification and test planning stages, but relax rigid architectural enforcement when functional correctness is the priority. (iii) \textit{Interface conformance check}: include a lightweight verification step to ensure output signatures, data types, and edge-case handling remain aligned with the defined contract. (iv) \textit{Model-stage alignment}: select models that respond positively to structured inputs, and use role-specific exemplars or context windows to prevent over-specification during design and validation phases.

\phead{For Researchers and Benchmark Designers.} Class-level generation shifts errors from low-level slips (e.g., missing code) toward higher-level reasoning and contract violations (semantic failures, return/type/format mismatches), revealing that stricter process \emph{changes} the failure profile rather than uniformly improving it. This has two consequences.
\emph{For researchers:} (i) Explore \textit{adaptive/conditional pipelines} that enable or skip stages only when they raise correctness; (ii) add \textit{cross-stage memory and reconciliation} so design/validation can’t silently override the specified contract; (iii) introduce \textit{conflict-resolution} (a meta-agent) that prefers benchmark contracts when spec vs. design disagree.
\emph{For benchmark designers:} (i) Publish explicit \textit{contract policies} (input ranges, error handling, formatting/typing rules) to reduce “reasonable but wrong” over-validation; (ii) report \textit{dual scores}: correctness (Pass@k) and design/cleanliness (e.g., maintainability/consistency metrics) so process benefits aren’t invisible; (iii) add \textit{spec-faithfulness checks} (API/signature/return-format conformance) and tag cases where extra policy is allowed vs. penalized; (iv) include \textit{ambiguity labels} or test rationales so agents learn when \emph{not} to tighten behaviour.

\subsection{Threats to Validity}
\phead{Internal Validity.}
We performed qualitative analysis to identify failure-inducing factors for RQ3. However, such analysis is inherently subjective and may not always reveal the exact cause of errors. Future work could incorporate LLM-based evaluators to diagnose error sources and reduce manual effort \cite{li2024llms}.
Minor inconsistencies may also arise in how errors are categorized across models. To minimize this, we randomly sampled generated outputs, applied consistent labelling, and reviewed all samples within the same time frame to reduce bias.

\phead{External Validity.}
A limitation of this work is the use of Pass@1 as the primary measure of functional correctness. Although Pass@1 evaluates only the first generated output and can be volatile, it remains a widely adopted and interpretable metric in LLM-based code generation research, used by Codex~\cite{chen2021evaluating}, AlphaCode~\cite{Li2022competition}, and CodeT5+~\cite{wang2023codet5}. This choice allows for consistent comparison across workflows and models.

\phead{Construct Validity.}
Our Multi-agent Waterfall workflow approximates its real-world counterpart, with LLMs playing the roles typically performed by humans. Different versions of the Waterfall model and varying phase combinations could affect results.
For test generation, we used LLM agents to produce unit-test suites, following prior studies that employ multi-agent LLM workflows for software development such as ChatDev \cite{qian2023chatdev}, MetaGPT \cite{hong2024metagpt}, and SOEN-101 \cite{lin2024soen}. These works demonstrate that LLM agents can generate runnable tests by interpreting code behaviour, compiler feedback, and execution traces, which we adopt in our setup. Future work could incorporate automated test-generation tools such as Pynguin \cite{lukasczyk2022pynguin} or explore hybrid LLM plus search-based methods \cite{broide2025evogpt} to further improve coverage and fault-detection capability \cite{make7030097}.
\section{Conclusion}
In this paper, we have evaluated multi-agent LLM workflows within the Waterfall model for class-level code generation using ClassEval. The experiments show that Waterfall workflows do not consistently improve functional correctness but yield cleaner, more maintainable code with fewer structural issues such as missing methods. Runtime and logical errors increase slightly, often due to stricter validation, semantic failures, and incorrect returns. Among development stages, Testing has the biggest impact, improving verification but reducing maintainability, while Requirement and Design play minor roles. Overall, the complete Waterfall workflow produces structured and readable code but limits flexibility and correctness, revealing a trade-off between disciplined process control and adaptive reasoning in multi-agent LLM systems.

\bibliographystyle{ACM-Reference-Format}
\bibliography{references}

\end{document}